{
\documentclass{article}
 \relax
\setlength{\oddsidemargin}{.25in}
\setlength{\textwidth}{6in}
\setlength{\topmargin}{.25in}
\setlength{\textheight}{7.5in}
\usepackage{graphics}

\begin{document}

\title{A WORMHOLE-GENERATED PHYSICAL UNIVERSE
}         
\author{A. L. Choudhury\footnote{Department of Physical Science, Elizabeth City State University,Elizabeth City, NC 27909} and Hemant Pendharkar\footnote{ Department of Computer Science and Mathematics, Elizabeth City State University, N C27909 }}      
\date{April 12, 2001}          
\maketitle

\begin{abstract}
   {   We constructed a model where the central core of the universe is a modified Gidding-Strominger wormhole and surrounding the core is a Robertson-Walker Universe with k=0. They are separated by a thin wall which does not allow the content of the inner core to travel to the outer universe. But this wall allows the pressure of the inner core to be transferred to the outer physical universe. Assuming that the fluid density of the physical universe is practically independent of time, we have calculated the Hubble constant and the deacceleration parameter, $q_0$, of the physical universe at the present time. The Hubble constant comes out to be positive, whereas $q_0$ becomes negative. The negative signature of this deacceleration parameter conforms to  present experimental data. }
\end{abstract}
\section{Introduction}
   {$\;$$\;$     In a recent paper Choudhury [1] has shown that a wormhole generated in a modified Gidding-Strominger model [2, 3] can possess a positive Hubble constant H0 and a negative deacceleration parameter $q_0$. Thus it shows that the wormhole scale factor has the right property to explain the current experimental results indicated in a report of Bahcall et al [4]. Picon et al [5] had to introduce complicated quintessential fields in order to explain experimental data. Their calculation turns out to be parameter dependent.

    We want to explore in this paper a new mechanism to generate a negative deacceleration parameter. We already know that the modified Gidding-Strominger model, which has a wormhole imbedded in the solution, can generate a positive Hubble parameter and a negative deacceleration parameter. These quantities can be calculated from the negative solution of the scale function a(t). In deriving these parameters we have used a conjecture originally introduced by Choudhury [6], which says that the wormhole universe expands according to the adiabatic gas law.

    We now modify the picture by introducing an extended Gidding-Strominger wormhole in the core of our physical universe and surrounding it with a Robertson-Walker space. We assume that they are separated by a thin wall which forbids a mixture of our physical universe with the content of the wormhole. However, any pressure generated in the wormhole can be transmitted to our physical universe by Pascal's law of fluid pressure transfer. We also assume in our preliminary investigation that the Robertson-Walker k-value to be zero. We would, however, leave the study of a non-zero value of k for future research activity.

    We shall assume that the Robertson-Walker space [7] contains a fluid with a constant density with respect to the change of R and a negligible self-pressure. The rationale of the constant density of the fluid can be justified as follows: since the Robertson-Walker physical universe is expanding and simultaneously the wormhole core is growing in size, the actual size of our physical universe can be assumed in the lowest order approximation to have the same size. Moreover the content of the matter does not change. As a consequence we can assume that the universe has a constant density as a function of the scale factor of the physical universe. However,  we  admit  that this assumption might not be rigidly true.

    Under those assumptions we then calculate the Hubble parameter and the deacceleration parameter for the physical universe. In section 2, we define the model of the inner core, the wormhole. This is nothing but a modified Gidding-Strominger model described by Choudhury [1]. We stress on the fact that the scale factor a(t) has two signatures.

    In section 3, we introduce our conjecture that the wormhole behaves like a gas satisfying the adiabatic gas law. The pressure of the wormhole is deduced in this section. In section 4, we describe briefly the Robertson-Walker universe with a scale factor R and show the relationship between R and the wormhole pressure. For a space k=0, we then calculate in section 5, the Hubble constant and deacceleration parameter $q_0$, both of which depends on the present time $t_0$. By adjusting the adiabatic gas constant g, we show that $H_0$ is positive and $q_0$ negative. In section 6, we discuss the result.}
\section{The Wormhole Core}
{$\;$$\;$The wormhole, which we intend to put in the core of the real expanding Robertson-Walker universe, is generated by a modified Gidding-Strominger action. While dealing with the inner core, we start in Euclidean space and get the solution of the wormhole scale function. We then switch over to Lorentz space-time by substituting t with it. Naturally we assume that such transformation can be done because the solution is analytic in a complex t-plane. For the core the action in the Euclidean space is given as follows:
\begin {equation}
S_E=\int {{d^4}x {{L_G}^c}(x)}+\int {{d^4}x {{L_{SA}}^c}(x)}=S_G+S_A,
\end{equation}
{where}
\begin {equation}
{L_G}^c=\frac{{\surd g^c}{R(g^c)}} {2{\kappa}^2},
\end{equation}
{and}
\begin{equation}
{{L_{SA}}^c} (x)={\surd {g^c}}[{\frac1 2}(\bigtriangledown \Phi)^2+{g_p}^2 {\Phi}^2 Exp(\beta {\Phi}^2)]{H_{\mu\nu\rho}}^2.
\end{equation}
{In the above expression and as well as in the subsequent formulae the suffix c stands for the core. Following Gidding and Strominger, we introduce}
\begin{equation}
H_{\mu\nu\rho}=\frac n {{g_p}^2 {a^3 (t)}} \varepsilon_{\mu\nu\rho},
\end{equation}
{from which we get}
\begin{equation}
H^2=\frac{6D} {{g_p}^2 {a^6}},
\end{equation}
{with}
\begin{equation}
D=\frac{n^2} {{g_p}^2}
\end{equation}
{The space-time interval in Euclidean space is given by}
\begin {equation}
ds^2=dt^2+a^2(t)(d\chi^2+sin^2 \chi d\theta^2+sin^2 \chi sin^2 \phi d\phi^2).
\end{equation}
{The variation of $g_{\mu\nu}^c$ in the core leads to the following equation}
\begin{equation}
G_{\mu\nu}^c=R_{\mu\nu}-{\frac1 2}g_{\mu\nu}^c R^c=\kappa^2[\nabla_\mu \Phi \nabla_\nu \Phi-({\frac1 2}(\nabla\Phi)^2) g_{\mu\nu}^c+{g_p}^2 Exp(\beta {\Phi}^2({H_\mu}^\alpha\gamma H_\nu\alpha\gamma- {\frac1 6} g_{\mu\nu}^c H_{\alpha\beta\gamma} ^2)]
\end{equation}
{For  the $\Phi$-variation inside the core the equation of motion of $\Phi$ yields  }
\begin{equation}
{\nabla}^2 \Phi-2{{g_p}^2\beta}\Phi{Exp(\beta\Phi^2)}{H_{\alpha\beta\rho}}^2=0.
\end{equation}
{ The Hamiltonian constraint yields}
\begin{equation}
(\frac1 a \frac{da} {dt})^2-\frac1 {a^2}=\frac{\kappa^2} 3 [{\frac1 2}(\frac{d\Phi} {dt})^2-6Exp(\beta\Phi^2) \frac{n^2} {g_p^2 a^6}].
\end{equation}
{The dynamical equation yields }
\begin{equation}
{\frac d {dt}} ({\frac1 a }{\frac{da} {dt}})+\frac1 {a^2}=-\kappa^2[(\frac{d\Phi} {dt})^2-12 Exp(\beta\Phi^2) {\frac{n^2} {g_p^2 a^6}}].
\end{equation}
{ Introducing a new variable $\tau$ defined by the relation}
\begin{equation}
d\tau=a^{-3} dt,
\end{equation}
{the Eq.(10) reduces to}
\begin{equation}
\frac{d^2 \Phi} {d\tau^2}-2D\beta Exp(\beta\Phi^2)=0.
\end{equation}
{The Eq.(13) can be converted into an integral equation given by }
\begin{equation}
\Phi(\tau)=\int_{0}^{t}\surd(2DExp(\beta\Phi^2(\tau\prime)-C_o)d\tau\prime,
\end{equation}
{where $C_o$ is an integration constant. Using this form of $\Phi(t)$, the Eqs. (10) and (11) can be combined into the following form: }
\begin{equation}
{(\frac1 a }{\frac{da} {dt}})^2-a^4+{a_c}^4=0,
\end{equation}
{where}
\begin{equation}
a_c=\surd(\frac{\kappa^2C_o} 2)
\end{equation}
{The solution of the Eqn.(15) has been obtained by Gidding and Strominger and is given by}
\begin{equation}
a^2(\tau)={a_c}^2\surd(sec(2{a_c}^2\tau)
\end{equation}
{The scale factor can now be of two possible forms }
\begin{equation}
a(\tau)=\pm a_c (sec(2{a_c}^2\tau)^{\frac1 4},
\end{equation}
{specified by the signs. Naturally both solutions stand on equal footing.}
\section{Pressure Generated By Wormhole}
{$\;$$\;$     We make now an assumption that the scalar fields are  an ensemble of particles expanding with the wormhole. We introduce here a classical picture assumed by Choudhury [1] that the gas is expanding adiabatically, following the  thermodynamical gas law}
\begin{equation}
P_w V_w ^\gamma=Constant=B_1,
\end{equation}
{where $\gamma$  is a constant. The volume of the wormhole can be shown to be }
\begin{equation}
V_w=2\pi^2a^3(\tau).
\end{equation}
{Combining Eqs.(19) and (20) we find}
\begin{equation}
P_w=B_1 V_w^{-\gamma}=Ba_{-3\gamma},
\end{equation}
{where we have used the abbreviation}
\begin{equation}
B=B_1(2\pi)^{-\gamma}
\end{equation}
{Inserting the value of $a(\tau)$ from Eq.(18) in Eq.(21) we get }
\begin{equation}
P_w=(\pm 1)^{-3\gamma}B{a_c}^{-3\gamma}[sec(2{a_c}^2 \tau)]^{-(3\gamma/4)}
\end{equation}
{If we now switch from Euclidean space to Lorentz space, in the zeroeth 
approximation, we replace in Eq.(12) a with $a_c$, and t with it. As a consequence we get }
\begin{equation}
\tau=ia_c^{-3}dt
\end{equation}
{ Substituting this value of $\tau$ in Eq.(23) we get the value of pressure as}
\begin{equation}
P_w=(\pm 1)^{-3\gamma}B{a_c}^{-3\gamma}[cosh(\frac{2t} {a_c})]^{(3\gamma/4)}
\end{equation}
{We are at liberty to choose a scalar field gas with suitable $\gamma$. We choose the value $\gamma=5/3$. The rationale of this choice is offered in section 5. We also choose the second sign of the pressure because both signs are equally valid. We justify the selection by claiming that the interior of our system, the wormhole, is an entity unobservable in the physical world. Just like the Fadeev ghost states in field theory it is used to eliminate unphysical properties of the model. We have intentionally used an unobservable core to generate time-dependent pressure. We hence write }
\begin{equation}
P_w=-B{a_c}^{5}[cosh(\frac{2t} {a_c})]^{(5/4)}
\end{equation}
\section{ The Robertson-Walker Space With A Wormhole At The Core}
{$\;$$\;$We assume that the universe we live in, is a Robertson -Walker space and  has a core of radius a which is the scale factor of the modified Gidding-Strominger wormhole. For the time being we assume that the wall between the wormhole and the physical universe does not allow material transfer. Thus the content of the wormhole matter and that of our universe do not mix. The only way they influence each other is through thermodynamic pressure. We assume that Pascal's law of fluid pressure holds. Consequently the pressure can be transferred undiminished into each other through the wall separating them. Of course, this assumption is purely hypothetical, and needs to be confirmed by more detailed thermodynamic calculation. However at this stage we assume the validity of the law and proceed to study the consequence of  this hypothesis. }
{$\;$$\;$     As usual our physical universe starts with a scale factor R(t), with metric given by the usual Robertson-Walker interval given by}
\begin{equation}
d \tau^2 = -d t^2 + {R^2}(t)[\frac{dr^2} {1-kr^2} + {r^2}d \theta^2 + {r^2}{sin^2}{\theta} d\phi^2]
\end{equation}
{ where we have chosen c=1. The Einstein equation is as follows:}
\begin{equation}
G_{\mu\nu}=R_{\mu\nu}-{\frac1 4}g_{\mu\nu} R = -{\frac{8 \pi G} 3}S_{\mu \nu} = - {\frac{8 \pi G} 3} (T_{\mu\nu}- {\frac1 4}g_{\mu\nu} {T^{\lambda}}_\lambda).
\end{equation}
{We assume that the energy momentum tensor has the perfect fluid form}
\begin{equation}
T_{\mu\nu}= P_T g_{\mu\nu}+(P_T +\rho) U_\mu U_\nu
\end{equation}
{where $P_T$, the total pressure and $\rho$, the density are  functions of t alone. The only non-vanishing component of U is the $U_t=1$. Following Weinberg[7] we find that the time component turns out to be }
\begin{equation}
3 \frac{d^2 R} {dt^2} = -{\frac {4 \pi G} 3}(\rho + 3 P_T )R ,
\end{equation}
{ and the space-space component}
\begin{equation}
R  \frac{d^2 R} {dt^2}+ 2 (\frac{dR} {dt})^2 + 2 k = {\frac {4 \pi G} 3}(\rho - P_T )R^2 .
\end{equation}
{ In Eq.(30) $P_T$ stands for  total pressure given by }
\begin{equation}
P_T = P_F + P_W
\end{equation}
{ where $P_F$ is the usual fluid pressure and $P_W$ is the wormhole pressure (Eq.(16)) which has been transferred to this point.}
{$\;$$\;$The equation of the energy conservation yields: }
\begin{equation}
{\frac{dP_T} {dt}} R^3= {\frac{d} {dt}} {[R^3 (\rho + P_T)]}.
\end{equation}
{Equivalently we can write}
\begin{equation}
{\frac{d} {dR}}( \rho R^3)= -P_T R .
\end{equation}
{ We now assume that the density $\rho$ of the physical universe in the first approximation is constant as a function of R. The justification of this assumption lies in the fact that when the wormhole core expands the physical universe, although expanding, is loosing some volume to make room for the expansion  of the wormhole. As a result the density can be imagined to stay the same. Hence we can conjecture that the r would not depend on time and the result does not depend on R(t). From Eq.(34) we can conclude}
\begin{equation}
\rho R^2 = - P_T R^2.
\end{equation}
{ On the other hand, combining the Eqs.(30) and (31), we find the well known relation}
\begin{equation}
(\frac{dR} {dt})^2 + k = {\frac {8 \pi G} 3} \rho R^2.
\end{equation}
{ Using  Eq.(35) we can rewrite Eqn.(36) as}
\begin{equation}
(\frac{dR} {dt})^2 + k = -{\frac {8 \pi G} 3} P_T R^2.
\end{equation}
{In this paper we only study the case of k=0. We find that}
\begin{equation}
\frac{(\frac{dR} {dt})^2} {R^2} = -{\frac {8 \pi G} 3} P_T .
\end{equation} 
{Assuming further that the fluid pressure of the physical world is significantly smaller than the wormhole pressure, we can replace PT by PW of Eq.(30). Therefore we can write }
\begin{equation}
\frac{(\frac{dR} {dt})} {R} = -\pm \surd({\frac {8 \pi G} 3} P_T ).
\end{equation}
\section{Hubble Constant And Deacceleration Parameter}
{$\;$$\;$We can now determine the Hubble constant and the deacceleration parameter of the physical universe, which has a core of a modified Gidding-Strominger wormhole. To obtain them we use the following expansion of the scale function R (see for example: Weinberg[7]):}
\begin{equation}
R(t)= R(t_o) [ 1 + H_o (t-t_o) - {\frac1 2} q_o {H_o}^2(t-t_o)^2 + ...],
\end{equation}
{where}
\begin{equation}
H_o = \frac{{\frac{dR} {dt}} (t_o)} {R(t_o)}
\end{equation}
{and}
\begin{equation}
q_o = - {\frac{{{\frac{d^2R} {dt^2}} (t_o)} {R(t_o)}}  {({\frac{dR} {dt}})^2 (t_o)}}
\end{equation}
{In the above relations t0 indicates the present time. Using Eq.(39) and Eq.(26), we find }
\begin{equation}
{\frac{\frac{dR} {dt}} R}= \pm\surd[i^2 {\frac{8 \pi G} 3}(-1)^{-3\gamma} {a_c}^{-3\gamma}(cosh{\frac{2t} a_c})^{3\gamma/2}] .
\end{equation}
{To turn to a physical expanding universe, we choose the positive sign in Eq.(43). We can now convert Eq.(43) to the following form:}
\begin{equation}
{\frac{\frac{dR} {dt}} R}= i^{-3\gamma +1+4n} \surd( {\frac{8 \pi G} 3}) {a_c}^{-3\gamma}(cosh{\frac{2t} a_c})^{3\gamma/4} .
\end{equation}
{The constant n is an arbitrary positive integer. We now introduce a constrain on $\gamma$ by demanding }
\begin{equation}
3\gamma =4n + 1.
\end {equation}
{Since $\gamma$ indicates a property originating from the wormhole, we would like to demand that the unphysical core can only possess the adiabatic gas constant $\gamma$=.33, or 1.67, or 3, etc. As an outcome we get the modified Eq.(44) in the form}
\begin{equation}
H_o={\frac{\frac{dR} {dt}(t_o)} {R(t_o)}}=  \surd( {\frac{8 \pi G} 3}) {a_c}^{-3\gamma}(cosh{\frac{2t_o} a_c})^{3\gamma/4} .
\end{equation}
{We find that the Hubble constant of the physical universe depends on the present time $t_o$ and is positive.}
{$\;$$\;$    To obtain the deacceleration parameter we have to differentiate Eq.(37) with respect to t and get}
\begin{equation}
2R {\frac{dR} {dt}} = - {\frac{8 \pi G} 3}{\frac{d{P_W}} {dt}}R^2 -{\frac{8 \pi G} 3} {P_W} 2R {\frac{dR} {dt}}.                                                                                                                                    \end{equation}
{We can  show that the derivative of  $P_W$ with respect to t at $t=t_o$ comes out to be}
\begin{equation}
({\frac{dP_W} {dt}}) _{t = t_o} = -3 \gamma B {a_c}^{-(3 \gamma + 1)} [cosh(\frac{2t_o} {a_c})]^{(3\gamma/2-1)} sinh(\frac{2 t_o} {a_c}) .
\end{equation}
{As a result the deacceleration parameter turns out as }
\begin{equation}
q_o(t_o) = -{\frac{4 \pi G} 3}[{ 3 \gamma B } {a_c}^{-(3 \gamma + 1)} (cosh\frac{2t_o} {a_c})^{(3\gamma/2-1)} sinh\frac{2 t_o} {a_c} {H_o}^2(t_o) - {a_c}^{-3 \gamma} (cosh\frac{2t_o} {a_c})^{3\gamma/2} {H_o}_{-1} (t_o)] .
\end{equation}
{For large $t_o$ since}
\begin{equation}
cosh\frac{2t_o} {a_c}\approx {\frac1 2} Exp(\frac{2t_o} {a_c}),
\end{equation}
{$H_o (t_o)$ becomes}
\begin{equation}
H_o(t_o) \approx {\surd({\frac{2 \pi G B} 3})} {a_c}^{-3\gamma/2} Exp({\frac{2t_o} {a_c}}).
\end{equation}
{We also note that for large $t_o$}
\begin{equation}
{H_o}^{-1}(t_o)\rightarrow 0.
\end{equation}
{As a result the deacceleration parameter turns out to be}
\begin{equation}
q_o(t_o)\approx -(\frac{2 \pi G B} 3)^2 (\frac{3 \gamma} {2^{3 \gamma /2}}) {a_c}^{-6\gamma-1} Exp[\frac{(3\gamma +4) t_o} {a_c}] .
\end{equation}
{We note that the deacceleration parameter at the present time is a negative quantity. As a consequence our physical universe is in an acceleration mode at the present time. The current experimental data hint at such an outcome.}
\section{Concluding Remarks}
{$\;$$\;$We have started with a wormhole core according to the prescription of a modified Gidding-Strominger model and assumed that the core expands adiabatically. This expansion generates a pressure, which depends on the scale factor of the wormhole. Due to two possibilities of the signature of the scale factor we can choose the negative value to generate negative pressure. We should remember here that the wormhole by itself has no physical significance except by generating pressure on the real physical world. We also assumed here that the physical universe outside the wormhole is a Robertson-Walker space separated by a wall from the wormhole. The pressure, which is generated in the wormhole due to the adiabatic expansion, penetrates through the wall in the physical universe according to Pascal's law of fluid pressure transfer.}
{$\;$$\;$We have started with a wormhole core according to the prescription of a modified Gidding-Strominger model and assumed that the core expands adiabatically. This expansion generates a pressure, which depends on the scale factor of the wormhole. Due to two possibilities of the signature of the scale factor we can choose the negative value to generate negative pressure. We should remember here that the wormhole by itself has no physical significance except by generating pressure on the real physical world. We also assumed here that the physical universe outside the wormhole is a Robertson-Walker space separated by a wall from the wormhole. The pressure, which is generated in the wormhole due to the adiabatic expansion, penetrates through the wall in the physical universe according to Pascal's law of fluid pressure transfer.}

{ $\;$$\;$   We assume the density of the physical universe in the first approximation, to be a constant if considered as a function of the scale factor. Moreover we assume that the pressure in the fluid is dominated by this transferred fluid pressure. The rationale of the first assumption can be justified by noticing the fact that the expanding physical universe looses some volume due to the expansion of the wormhole core. As a result the volume of the physical universe would not change significantly. Hence in the first approximation, we can assume that the density of the physical world does not change with the change of R, justifying the assumption that the derivative of r with respect to R is zero. We also assumed k=0, to initiate the investigation.}

{$\;$$\;$    Under the above assumptions, we find that the present value of $H_o$ is positive and the deacceleration parameter is negative. For a large value $t_o$, we find $q_o$ depends exponentially on $t_o$. Since present experimental results indicate that the deacceleration parameter appears to be negative, we claim that our model can be an alternative description of our physical world.}

{$\;$$\;$    Although we have used the negative alternative of the wormhole scale function a(t), it is still conceivable that the positive value of the scale function might have negative $q_o$ under suitable adjustment of the parameters introduced in the model. For the non-zero values of k in the Robertson-Walker space we are also contemplating to study the properties of H0 and q0 along this line. Work is already in progress.}
\section{References}
\begin{enumerate}
\item { A. L. Choudhury, Hadronic J., 23, 581 (2000).}
\item { S. B. Giddings and A.  Strominger , Nucl. Phys. B 307, 854 (1988).}
\item { D. H. Coule and K. Maeda, Class. Quant. Grav. 7, 955 (1990).}
\item { N. Bahcall, J. P. Ostriker, S. Perlmutter, and P. J. Steinhardt, Science, 284,1481 (1999).}
\item { C. Aremendariz Picon, V. Mukhanov and Paul J. Steinhardt: Essentials of k-Essence. ArXiv:astro-ph/0006373 (2000).}
\item { A. L. Choudhury, Hadronic J. Suppl. 13, 395 (1998).}
\item { S. Weinberg: Gravitational Cosmology, John Wiley and Sons, 340 (1972).}
\end{enumerate}    


\end{document}